\documentclass[twocolumn,showpacs,aps,prl,superscriptaddress,letterpaper]{revtex4}
\usepackage{graphicx}
\usepackage{dcolumn}
\usepackage{amsmath}
\usepackage{epsfig}
\long\def\inst#1{\par\nobreak\kern 4pt\nobreak
    {\itshape #1}\par\vskip 10pt plus 3pt minus 3pt}
\RequirePackage{xspace}
\usepackage{relsize}
\def\qqbar {\ensuremath{q\overline q}\xspace}
\def\babar{\mbox{\slshape B\kern-0.1em{\smaller A}\kern-0.1em
    B\kern-0.1em{\smaller A\kern-0.2em R}}}
\def\Dbar    {\kern 0.18em\overline{\kern -0.18em D}{}\xspace}
\def\Bbar    {\kern 0.18em\overline{\kern -0.18em B}{}\xspace}

\def\BB      {\ensuremath{B\Bbar}\xspace} 
\def\Bz      {\ensuremath{B^0}\xspace}
\def\Bzb     {\ensuremath{\Bbar^0}\xspace}
\def\BzBzb   {\ensuremath{\Bz {\kern -0.16em \Bzb}}\xspace}
\def\Bu      {\ensuremath{B^+}\xspace}
\def\Bub     {\ensuremath{B^-}\xspace}

\def\BpBm    {\ensuremath{\Bu {\kern -0.16em \Bub}}\xspace}

\newcommand{\optbar}[1]{\shortstack{{\tiny (\rule[.4ex]{1em}{.1mm})}
  \\ [-.7ex] $#1$}}
\def\BorBbar    {\kern 0.18em\optbar{\kern -0.18em B}{}\xspace}
\def\DorDbar    {\kern 0.18em\optbar{\kern -0.18em D}{}\xspace}
\def\KorKbar    {\kern 0.18em\optbar{\kern -0.18em K}{}\xspace}
\def\CP                {\ensuremath{C\!P}\xspace}
\def\pep2{PEP-II}
\mathchardef\Upsilon="7107
\def\Y#1S{\ensuremath{\Upsilon{(#1S)}}\xspace}

\def\FourS {\Y4S}

\newcommand{\BABARPubYear}     {03}
\newcommand{\BABARPubNumber}  {040}
\newcommand{\SLACPubNumber} {10217}

\begin{document}

\begin{flushleft}
\babar-PUB-\BABARPubYear/\BABARPubNumber\\
SLAC-PUB-\SLACPubNumber
\\[5mm]
\end{flushleft}

\title{
\large \bfseries \boldmath
Observation of the Decay $B^0\to\rho^+\rho^-$ and\\ 
Measurement of the Branching Fraction and Polarization 
}

%
\author{B.~Aubert}
\author{R.~Barate}
\author{D.~Boutigny}
\author{F.~Couderc}
\author{J.-M.~Gaillard}
\author{A.~Hicheur}
\author{Y.~Karyotakis}
\author{J.~P.~Lees}
\author{V.~Tisserand}
\author{A.~Zghiche}
\affiliation{Laboratoire de Physique des Particules, F-74941 Annecy-le-Vieux, France }
% ...
\collaboration{The \babar\ Collaboration}
\noaffiliation


\date{November 6, 2003}


\begin{abstract}
We have observed the rare decay $B^0\to\rho^+\rho^-$
in a sample of 89 million $\BB$ pairs recorded 
with the $\babar$ detector.
The number of observed events is $88^{+23}_{-21}\pm 9$, 
with a significance of 5.1
standard deviations with systematic uncertainties included.
The branching fraction and the longitudinal 
polarization are measured to be
${\cal B}(B^0\to\rho^+\rho^-)=(25^{+7+5}_{-6-6})\times 10^{-6}$
and ${\Gamma_L}/\Gamma=0.98^{+0.02}_{-0.08}\pm 0.03$,
respectively.
\end{abstract}


\maketitle


Charmless $B$-meson decays provide an opportunity to measure 
the angles of the unitary triangles constructed from the elements of the 
Cabibbo-Kobayashi-Maskawa (CKM) quark-mixing matrix~\cite{Kobayashi}.
There has been interest in the study
of $B\to\pi\pi$ and $\rho\pi$ decays, where the time-dependent
$\CP$-violating asymmetries are related to the CKM angle
$\alpha\equiv {\rm arg}\,[\, - V^{ }_{td}V^*_{tb}\,/
\,V^{}_{ud}V^*_{ub}\,]$,
and interference between tree and
loop (penguin) amplitudes could give rise to direct $\CP$ violation.
The decay $B^0\to\rho^+\rho^-$ is another promising 
mode for $\CP$-violation studies and has the advantage
of a larger expected decay rate and smaller uncertainty in penguin
contributions. The measurements of the 
amplitudes in $B$ decays to two vector particles provide
additional tests of
theoretical calculations~\cite{bvv1, bvv2, bvv3}.

\begin{figure}
\begin{center}
\setlength{\epsfxsize}{1.0\linewidth}\leavevmode\epsfbox{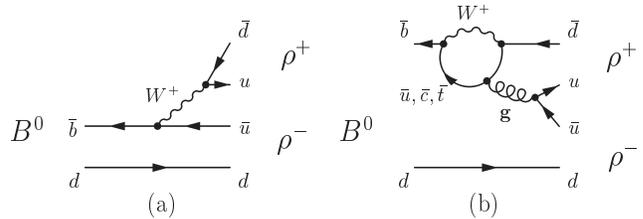}
\caption{
Diagrams describing the decay $B^0\to\rho^+\rho^-$:
(a) dominant tree diagram, (b) gluonic penguin diagram.
}
\label{fig:Diagram} 
\end{center}
\end{figure}

The decay $B^0\to\rho^+\rho^-$ is expected to proceed through 
the tree-level $b\to u$ transition and through
CKM-suppressed $b\to d$ penguin transitions, as illustrated in 
Fig.~\ref{fig:Diagram}~\cite{bvv3, chargeconj}.
The extraction of $\alpha$ from measurements made with
this decay requires an understanding of the 
contributing amplitudes. It also requires proper 
accounting for $\CP$-even (S- and D-wave) and $\CP$-odd 
(P-wave) components in the decay amplitude.
The recent limit on the $B^0\to\rho^0\rho^0$ 
decay rate~\cite{babarVV} and the measurements of the
$B^+\to\rho^+\rho^0$ branching fraction~\cite{babarVV, belle} 
place experimental limits on the 
contribution of penguin amplitudes.
Measurements of the longitudinal polarization, 
defined as the ratio between the longitudinal and total decay 
rates $f_L\equiv{\Gamma_L}/{\Gamma}$~\cite{bvv1}, 
in the $B^+\to\rho^+\rho^0$ decay provide evidence that the 
$\CP$-even component dominates in 
$B\to\rho\rho$ decays~\cite{babarVV, belle}.

In this paper we report the observation of the 
$B^0\to\rho^+\rho^-$ decay mode and measurements
of its branching fraction and the amount of longitudinal polarization
in the decay. We also make a quantitative estimate 
of penguin contributions in this decay using our earlier 
measurements in isospin-related $B\to\rho\rho$ modes.

We use data collected with the 
\babar\ detector~\cite{babar} at the \pep2 asymmetric-energy 
$e^+e^-$ storage ring.
These data represent an integrated luminosity 
of 81.9~fb$^{-1}$ at the $e^+e^-$ center-of-mass (CM) energy 
of the $\FourS$ resonance ($\sqrt{s}=10.58$~GeV, on-resonance),
corresponding to 88.9 million $\BB$ pairs, 
and 9.6~fb$^{-1}$ approximately 40~MeV below this energy 
(off-resonance). 

Charged-particle momenta are measured in a tracking system 
consisting of a five-layer double-sided silicon vertex tracker 
(SVT) and a 40-layer central drift chamber (DCH), 
both situated in a 1.5-T axial magnetic field. 
\babar\ achieves an impact parameter resolution
of about 40~$\mu$m for the high-momentum charged particles
from the $B$ decay, allowing the precise 
determination of decay vertices.
The tracking system covers 92\% of the solid angle in the CM frame.

Charged-particle identification is provided by 
measurements of energy loss 
(${\rm d}E/{\rm d}x$) in the tracking devices (SVT and DCH) and
by an internally reflecting ring-imaging Cherenkov detector 
(DIRC). A $K$-$\pi$ separation of 
better than four standard deviations ($\sigma$) is achieved for 
momenta below 3~GeV, decreasing to 2.5~$\sigma$ at the highest 
momenta in the $B$ decay final states.
Photons are detected by a CsI(Tl) electromagnetic calorimeter
(EMC). The EMC provides good
energy and angular resolution for detection of photons 
with energy in the range 20~MeV to 4~GeV.
The energy and angular resolutions are $3\%$ and 4 mrad, 
respectively, for a 1~GeV photon.


Hadronic events are selected based on track multiplicity and 
event topology. We fully reconstruct $B^0\to\rho^+\rho^-$ 
candidates from the decay products of the
$\rho^\pm\to\pi^\pm\pi^0$ and $\pi^0\rightarrow \gamma\gamma$
decays. Charged-track candidates are required to originate 
from the interaction point, have at least 12 DCH hits 
and have a minimum transverse momentum of 0.1~GeV. 
Charged-pion tracks are distinguished from kaon and proton tracks 
with a likelihood ratio that includes 
${\rm d}E/{\rm d}x$ information from the SVT and DCH, and,
for momenta above 0.7~GeV, 
the Cherenkov angle and number of photons measured by the DIRC.
Charged pions are distinguished from electrons primarily on 
the basis of their EMC shower energy and spatial profile.

\begin{figure}
\begin{center}
\setlength{\epsfxsize}{1.0\linewidth}\leavevmode\epsfbox{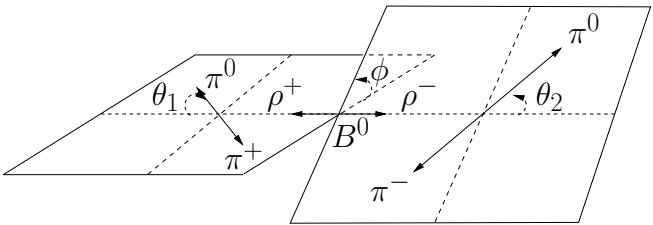}
\caption{
Definition of helicity angles $\theta_1$, $\theta_2$, and
$\phi$, for the decay ${B^0\to\rho^+\rho^-}$. The $\pi^\pm\pi^0$ 
final states are shown in the $\rho^\pm$ rest frames.
}
\label{fig:helangles} 
\end{center}
\end{figure}

We reconstruct $\pi^0$ mesons from pairs of photons.
Photon candidates are required to have
a minimum energy of 30 MeV, have a shower shape consistent
with the photon hypothesis, and not be matched to a track.
The typical experimental resolution for the measured
$\pi^0$ mass is 7~MeV. 
We require $\pi^0$ candidates to have an invariant mass
within 15~MeV of the true $\pi^0$ mass.
The invariant mass of the $\rho^\pm$ candidate ($m_{\pi^\pm\pi^0}$) 
is required to be in the range 0.52 to 1.02~GeV.
The helicity angles $\theta_{1}$ and $\theta_{2}$
of $\rho^+$ and $\rho^-$
are defined as the angles between the $\pi^0$ direction
and the direction opposite the $B$ in each $\rho$ rest frame
as shown in Fig.~\ref{fig:helangles}.
The helicity angles are restricted
to the region $-0.75\le\cos\theta_{1,2}\le0.95$
to suppress combinatorial background and 
reduce acceptance uncertainties due to 
low-momentum pion reconstruction.

The $B$ meson candidates are identified from two nearly 
independent kinematic observables \cite{babar},
the beam energy-substituted mass $m_{\rm{ES}} =$ 
$[{ (s/2 + \mathbf{p}_i \cdot \mathbf{p}_B)^2 / E_i^2 - 
\mathbf{p}_B^{\,2} }]^{1/2}$ and the energy difference
$\Delta E = (E_i E_B - \mathbf{p}_i 
\cdot \mathbf{p}_B - s/2)/\sqrt{s}\,$,
where $(E_i,\mathbf{p}_i)$ is the $e^+e^-$ initial state 
four-momentum, and $(E_B,\mathbf{p}_B)$
is the four-momentum of the reconstructed $B$ candidate,
all defined in the laboratory frame.
For signal events, the $m_{\rm{ES}}$ distribution 
peaks at the $B$ mass and the $\Delta E$ distribution 
peaks near zero.
Our selection requires $m_{\rm{ES}}>5.2$~GeV
and $|\Delta E|<0.2$~GeV, while the signal resolution
is roughly 3~MeV and 50~MeV, respectively. 
The sideband regions are defined as 
$5.2~{\rm GeV}<m_{\rm{ES}}<5.27$~GeV
or $0.1~{\rm GeV}<|\Delta E|<0.2$~GeV.

To reject the dominant continuum background 
(from $e^+e^-\to\qqbar$ events, $q = u, d, s, c$),
we require $|\cos\theta_T| < 0.8$, where $\theta_T$ 
is the angle between the thrust axis of the $B$ candidate 
and the thrust axis
of the rest of the tracks and photon candidates in
the event, calculated in the CM frame. 
The distribution of $|\cos{\theta_T}|$ is 
sharply peaked near 1.0 for jet-like events originating 
from $\qqbar$ pairs and nearly uniform for the 
isotropic decays of the $B$ meson.
A Fisher discriminant (${\cal F}$) combines 11 observables: 
the polar angle of the $B$ momentum vector and
the polar angle of the $B$-candidate thrust axis,
both calculated with respect to the beam axis in the CM frame,
and the scalar sum of the CM momenta of charged particles
and photons (excluding particles from the $B$ candidate)
entering nine coaxial angular intervals of 10$^\circ$
around the $B$-candidate thrust axis~\cite{CLEO-fisher}.

The selected sample contains 54,042 events most of which 
populate sidebands of the observables.
Background from other $B$ decays is estimated with
Monte Carlo (MC) simulation~\cite{geant}; it
contributes 5\% of the events in the selected sample.
This background component, arising mainly from $b\to c$ transitions, 
is explicitly included in the fit described below.


We use an unbinned, extended maximum-likelihood fit to
extract simultaneously the signal yield and polarization. 
There are three event categories $j$: signal, 
continuum~$\qqbar$, and $\BB$ combinatorial background.
The likelihood for each $B^0\to\rho^+\rho^-$ candidate~$i$
is defined as
\begin{equation}
{\cal L}_i = \sum_{j=1}^{3} n_{j}\, 
{\cal P}_{j}(\vec{x}_{i};\vec{\beta}) ,
\label{eq:likelev}
\end{equation}
where each of the ${\cal P}_{j}(\vec{x}_{i};\vec{\beta})$ is 
the probability density function (PDF) for seven observables
$\vec{x}_{i}$ ($m_{\rm{ES}}$, $\Delta E$, ${\cal F}$, 
$m_{\pi^{+}\pi^0}$, $m_{\pi^{-}\pi^0}$,
$\theta_{\rm 1}$, $\theta_{\rm 2}$) and is described by
the PDF parameters $\vec{\beta}$.
The event yields $n_{j}$ for each category $j$ are free 
parameters in the fit.
We allow for multiple candidates in a given event by assigning 
to each selected candidate a weight of $1/N_i\,$, 
where $N_i$ is the number of candidates in that event.
The average number of candidates per event is 1.27.
MC simulation shows that this procedure does not
introduce bias while providing a small statistical improvement 
over the random choice of a candidate in a given event.
The extended likelihood for a sample 
of $N_{\rm cand}$ candidates is 
\begin{equation}
{\cal L} = \exp\left(-\sum_{j=1}^{3} n_{j}\right)\, 
\prod_{i=1}^{N_{\rm cand}} 
\exp\left(\frac{\ln{\cal L}_i}{N_i}\right) .
\label{eq:likel}
\end{equation}

The correlations among the input observables $\vec{x}_{i}$
are found to be small for both the background ($<$5\%) 
and signal ($<$10\%), except for angular 
correlations in the signal.
The ${\cal P}_{j}(\vec{x}_{i};\vec{\beta})$,
for a given candidate $i$, is the product of PDFs
for each of the observables and a joint PDF for the  
helicity angles, which accounts for the angular correlations 
in the signal and for detector acceptance effects.
We integrate over the angle $\phi$ between the two 
decay planes shown in~Fig.~\ref{fig:helangles}, 
leaving a PDF that depends 
only on $\theta_{\rm 1}$, $\theta_{\rm 2}$,
and the unknown longitudinal 
polarization $f_L$.
The differential decay rate~\cite{bvv1} is
\begin{eqnarray}
{\frac{1}{\Gamma}} \  
{\frac{{\rm d}^2\Gamma}
{{\rm d}\cos \theta_1 \, {\rm d}\cos \theta_2}}
=~~~~~~~~~~~~~~~~~~ \cr
{\frac{9}{4}} \left \{ {\frac{1}{4}} (1 - f_L)
\sin^2 \theta_1 \sin^2 \theta_2 + f_L \cos^2 \theta_1 \cos^2 \theta_2 \right\} \ .
\label{eq:helicityintegr}
\end{eqnarray}

The PDF parameters $\vec{\beta}$, except for $f_L$,
are extracted from MC simulation and on-resonance 
$m_{\rm{ES}}$ and $\Delta E$ sidebands, and are fixed
in the fit.
The resolutions are adjusted by comparing data and simulation 
in calibration channels with similar kinematics and topology,
such as $B\rightarrow\Dbar\rho^+, \Dbar\pi^+$ with 
$\Dbar\rightarrow K^+\pi^-(\pi^0), K^0\pi^-(\pi^0), 
K^+\pi^-\pi^-, K^0\pi^-\pi^+$. 
To describe the signal distributions, we use Gaussian functions 
for the parameterization of the PDFs for $m_{\rm{ES}}$ and
$\Delta E$, and a relativistic P-wave Breit-Wigner distribution
for the $\rho^\pm$ resonance masses. The angular acceptance effects
are parameterized with empirical polynomial functions for each
helicity angle and are included in the joint helicity-angle PDF as 
a factor multiplying 
the ideal distribution in Eq.~(\ref{eq:helicityintegr}).

For the background PDFs, we use polynomials or, in the 
case of $m_{\rm{ES}}$, an empirical phase-space function~\cite{argus}.
In the background PDF  we incorporate a small linear 
correlation between the curvature $\xi$ of the phase-space function  
and the value of ${\cal F}$.
The background parameterizations for the $\rho^\pm$ candidate masses 
also include a resonant component to account for $\rho^\pm$ production.
The background helicity-angle distribution is 
also separated into contributions from combinatorial background
and from real $\rho^\pm$ mesons, both described by polynomials.
For both signal and background, the PDF for ${\cal F}$ 
is represented by a Gaussian distribution 
with different widths above and below the peak.

PDF parameters for the background from other $B$ decays 
are determined from MC simulation.
The contribution from charmless $B$ decays with similar 
topology (cross-feed modes) such as $B\to\rho\pi$, 
$\rho^0\rho^+$, $\rho K^*$, $a_1\pi$, and $a_1\rho$ 
is estimated with MC modeling and is fixed in the fit.
Each branching fraction for the cross-feed modes
is estimated to be in the range $(1\hbox{--}3)\times 10^{-5}$. 
The branching fractions for these and many other modes 
are taken from the most recent 
measurements~\cite{babarVV, belle, pdg} or
extrapolated from other results with a 
flavor-SU(3)-symmetry approximation. 

The selected  $B^0\to\rho^+\rho^-$ events fall into three categories.
MC simulation of events with longitudinal polarization shows that
roughly 30\% of the events contain only misreconstructed
candidates. Approximately 20\% of the events contain both
correctly and incorrectly reconstructed candidates.
The remainder contain only correct candidates.
Misreconstruction occurs when
at least one candidate photon in a $\pi^0$ 
candidate or one charged track in a $\rho$ candidate 
belongs to the decay products of the other $B$.  The distributions that
show peaks for correctly reconstructed events have substantial tails,
with large uncertainties in MC simulation,
when misreconstructed events are included.  These tails would reduce the power
of the distributions to discriminate between the background and the 
collection of correctly and incorrectly reconstructed events.  We
choose, therefore, to represent only the correctly reconstructed candidates
in the signal PDF. Misreconstructed candidates are predominantly
accommodated by the combinatorial background PDF.  Fitting to determine
the number of correctly reconstructed candidates has an efficiency
less than 100\% since some fraction of the events have both correctly
and incorrectly reconstructed candidates. Monte Carlo simulation finds 
this efficiency to be 87\%.

In this analysis, we do not include a fit component for 
other $B$ decays with the same final-state particles 
selected within the $\rho$ resonance mass window, 
such as nonresonant decays $B^0\to\pi^+\pi^-\pi^0\pi^0$
and $B^0\to\rho^\pm\pi^\mp\pi^0$.
The contribution of these decays to the fit results
is significantly suppressed 
by the selection requirements on the masses and by the mass
and helicity-angle information in the fit; 
they are examined in the context of mass and helicity-angle
distributions, as discussed below. 

The event yields $n_j$ and polarization $f_L$ are obtained by 
minimizing the quantity $\chi^2\equiv -2\ln{\cal L}$. 
The dependence of $\chi^2$ on a fit parameter
$n_j$ or $f_L$ is obtained with the other
fit parameters floating. Their values are constrained 
to the physical range $n_j\geq 0$ and $0\le f_L\le 1$.
Statistical uncertainties correspond to a unit
increase in $\chi^2$.
The statistical significance of the signal is defined as the 
square root of the change in $\chi^2$ when 
the number of signal events is constrained
to zero in the likelihood fit.


The results of our maximum-likelihood fits are summarized in 
Table~\ref{tab:results}. The statistical significance of the 
$B^0\to\rho^+\rho^-$ signal is 5.5~$\sigma$.
We find that the $\rho^\pm$ mesons in $B^0\to\rho^+\rho^-$
decays are almost fully longitudinally polarized. 
To compute the branching fraction, 
equal production rates for $\BzBzb$ and $\BpBm$ are assumed.
To check the stability of our results we refit, removing each 
observable from the fit in turn, and find consistent results.
The measured uncertainties in the number of fitted events
and the polarization, the statistical significance,
and the fit $\chi^2$ value are well 
reproduced with generated MC samples.

\begingroup
\begin{table}[btp]
\caption{Summary of the fit results;
$n_{\rm sig}$ is the fitted number of signal events, 
${\cal S}$ is the significance,
$f_L$ is the longitudinal polarization,
$\varepsilon$ denotes the reconstruction efficiency, and
${\cal B}$ is the branching fraction of
the $B^0\to\rho^+\rho^-$ decays.
The first uncertainty is statistical and the second systematic.
The efficiency ($\varepsilon$) and 
significance (${\cal S}$)
include systematic uncertainties, and
the significance without systematics
is given in parentheses.}
\label{tab:results}
\footnotesize
\begin{center}
\begin{ruledtabular}
\setlength{\extrarowheight}{1.5pt}
\begin{tabular}{cccc}
\vspace{-2.5mm}& &  & \\
 & Quantity & Measured Value  & \\
\vspace{-2.5mm}& &  & \\
\hline
\vspace{-2.5mm}& &  & \\
 & $n_{\rm sig}$  & $88^{+23}_{-21}\pm 9$  & \\
\vspace{-2.5mm}& &  & \\
 & ${\cal S}$ & 5.1 $\sigma$ (5.5 $\sigma$)  &  \\
\vspace{-2.5mm}& &  & \\
 & $f_L$ & $0.98^{~+0.02}_{~-0.08}\pm 0.03$  & \\
\vspace{-2.5mm}& &  & \\
 & $\varepsilon$ & $3.9_{-0.6}^{+0.9}~\%$  & \\
\vspace{-2.5mm}& &  & \\
 & ${\cal B}$ & $(25^{~+7~+5}_{~-6~-6})\times{10^{-6}}$  & \\
\vspace{-2.5mm}& &  & \\
\end{tabular}
\end{ruledtabular}
\end{center}
\end{table}
\endgroup

\begin{figure}
\setlength{\epsfxsize}{1.0\linewidth}\leavevmode\epsfbox{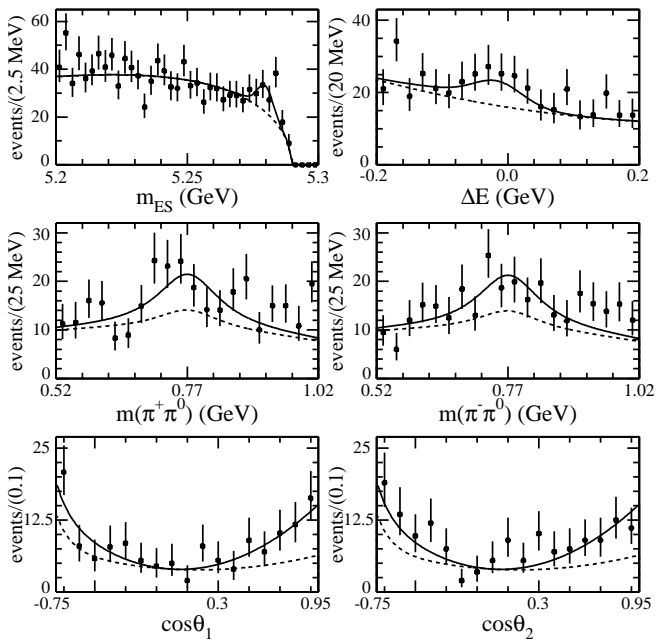}
\caption{
Projections onto the observables 
$m_{\rm{ES}}$, $\Delta E$,
$m_{\pi^+\pi^0}$, $m_{\pi^-\pi^0}$, 
$\cos\theta_1$, and $\cos\theta_2$ 
after a requirement on 
the signal-to-background probability ratio
${\cal P}_{\rm{sig}}/{\cal P}_{\rm{bkg}}$ with the 
PDF for the plotted observable excluded.
The points with error bars show the data, the solid (dashed) 
line shows the signal-plus-background 
(background only) PDF projection.
}
\label{fig:proj} 
\end{figure}

The projections of the fit input observables
are shown in Fig.~\ref{fig:proj}.
The projections are made after a requirement on 
the signal-to-background probability ratio
${\cal P}_{\rm{sig}}(\vec{x}_{i};\vec{\beta})/
{\cal P}_{\rm{bkg}}(\vec{x}_{i};\vec{\beta})$, where 
${\cal P}_{\rm{sig}}$ and ${\cal P}_{\rm{bkg}}$
are the signal and the dominant continuum background PDFs 
defined in Eq.~(\ref{eq:likelev}), but with the PDF for the 
plotted observable excluded. 
The points with error bars show the data with (40--60)$\%$
of signal retained, while the lines show the corresponding
PDF projections.

To check the sensitivity of our results to the 
presence of nonresonant $B^0\to\pi^+\pi^-\pi^0\pi^0$
and $B^0\to\rho^\pm\pi^\mp\pi^0$ decays, we
explicitly include a fit component for them, assuming 
a phase-space decay model. The selection requirements
alone suppress the $B\to 4\pi$ ($B\to\rho\pi\pi$) 
efficiency by two (one) orders of magnitude 
relative to $B^0\to\rho^+\rho^-$. 
The fit results with a nonresonant component 
indicate a potential $B\to\rho\pi\pi$ 
contribution of $(10\pm{10})\%$ (statistical uncertainty only)
of our nominal $B^0\to\rho^+\rho^-$ event yield 
in Table~\ref{tab:results}; 
interference effects between the resonant and nonresonant 
components were ignored in this fit. 
The hypothesis that all the signal is nonresonant  
$B\to 4\pi$ ($B\to\rho\pi\pi$) is excluded with 
$5.1$~$\sigma$ ($4.4$~$\sigma$) statistical significance. 
These results are consistent with our assumption that the
nonresonant contribution is negligible.

The systematic uncertainty in the fitted number
of signal events ($n_{\rm sig}$)
originates from the uncertainty in the 
cross-feed $B$-decay modeling, which was studied with 
MC generated samples and estimated to be half of the 
variation with cross-feed set to zero 
(3\% uncertainty in $n_{\rm sig}$).
Systematic uncertainties in the fit originate from assumptions 
about the PDF parameters.
Uncertainties in the PDF parameters arise from the limited
number of events in the background sideband data and signal 
control samples.
We vary them within their respective uncertainties,
and derive the associated systematic 
uncertainty on the event yield (9\%).
The signal remains statistically significant with
these variations ($5.1$~$\sigma$ including systematics).

The systematic uncertainties in the efficiency ($\varepsilon$)
are due to track finding 
(2\% for two tracks), particle identification (2\% for two tracks), 
and $\pi^0$ reconstruction (13\% for two $\pi^0$s). 
The fit efficiency is less than 100\% because of 
misreconstructed signal events. This has
an additional systematic uncertainty due to 
uncertainties in the modeling of misreconstructed events.
We account for this with a systematic uncertainty 
on the efficiency of 7\%, which is half of the inefficiency;
the fit efficiency cannot exceed 100\% 
and the frequency of multiple candidate 
selection is estimated in the $B$ decay control samples.
The reconstruction efficiency depends on the decay 
polarization. We calculate the efficiencies using 
the measured polarization and assign 
a systematic uncertainty (${}_{~-3}^{+17}\%$)
corresponding to the total polarization measurement 
uncertainty. Smaller systematic uncertainties arise
from event-selection criteria, MC statistics, 
and the number of produced $B$ mesons.

For the polarization measurement ($f_L$), 
we include systematic uncertainties from PDF variations 
that account for uncertainties in the detector acceptance,
estimated with MC, and background parameterizations.
This results in a total absolute uncertainty of 0.025.
The biases from the resolution in helicity-angle measurement 
and dilution due to the presence of the misreconstructed 
combinations are studied with MC simulation and 
give a systematic uncertainty of 0.02.


Observation of the $B^0\to\rho^+\rho^-$ decay
completes a first set of measurements of the isospin-related  
$B\to\rho\rho$ modes~\cite{babarVV, belle}.
The measured branching fraction is consistent with 
recent predicted values in the range 
$(18\hbox{--}35)\times 10^{-6}$~\cite{bvv3}
and the dominant longitudinal polarization implies 
a suppression of the transverse amplitude, 
which is expected to be suppressed
by a factor of $m_\rho/m_B$~\cite{bvv3}.
The rates of the $B^0\to\rho^+\rho^-$ and $B^+\to\rho^0\rho^+$ 
decays appear to be larger than 
the corresponding rates of $B\to\pi\pi$ decays~\cite{pdg}.
At the same time, the recent measurement 
of the $B^+\to\rho^0K^{*+}$ branching 
fraction~\cite{babarVV} does not show significant enhancement 
with respect to $B\to\pi K$ decays~\cite{pdg}, both of which
are expected to be dominated by $b\to s$ penguin diagrams.
We can use flavor SU(3) to relate $b\to s$ and 
$b\to d$ penguins analogous to 
Fig.~\ref{fig:Diagram}(b)~\cite{GronauRosner};
the measured branching fractions indicate that
the relative penguin contributions in the $B\to\rho\rho$ 
decays are smaller than in the $B\to\pi\pi$ case.

We make a more quantitative estimate of penguin contributions 
in $B\to\rho\rho$ decays using our previous measurements of 
$B^0\to\rho^0\rho^0$ and $B^+\to\rho^+\rho^0$ branching 
fractions and polarization~\cite{babarVV}. 
Since the tree contribution to the $B^0\to\rho^0\rho^0$ 
decay is color-suppressed, 
the decay rate is sensitive to the penguin diagram analogous to
Fig.~\ref{fig:Diagram}(b).
Using the earlier $\babar$ measurements~\cite{babarVV}, we obtain 
a 90\% confidence level (C.L.) upper limit on the ratio of the
longitudinal amplitudes $A_L$ in the $B\to\rho\rho$ decays:
\begin{eqnarray}
\frac{|{A_L}(B^0\to\rho^0\rho^0)|^2+|{A_L}(\Bbar^0\to\rho^0\rho^0)|^2}
     {2\times|{A_L}(B^+\to\rho^0\rho^+)|^2} = \cr\cr\cr
\frac{{\cal B}(B^0\to\rho^0\rho^0)\times f_L(B^0\to\rho^0\rho^0)}
     {{\cal B}(B^+\to\rho^0\rho^+)\times f_L(B^+\to\rho^0\rho^+)}
< 0.10 \, .
\label{eq:rhorhoratio}
\end{eqnarray}
In the above calculation
we conservatively assume that the $B^0\to\rho^0\rho^0$ decay
polarization is fully longitudinal ($f_L=1$),
use the average branching fraction measurements
for the $B$ and $\Bbar$ decays, and assume 
$|{A_L}(B^+\to\rho^0\rho^+)|=|{A_L}(B^-\to\rho^0\rho^-)|$
with only a tree-diagram contribution.
The limit in Eq.~(\ref{eq:rhorhoratio}) corresponds to
a $19^\circ$ uncertainty (at 90\% C.L.) on $\alpha$ 
due to penguin contributions in the time-dependent 
measurements with longitudinally-polarized
$B^0\to\rho^+\rho^-$ decays, assuming isospin relations 
analogous to those discussed in the context of 
$B\to\pi\pi$~\cite{grossmanquinn} 
and neglecting the nonresonant and $I=1$ isospin
contributions~\cite{falk}.

In summary, we have observed the decay $B^0\to\rho^+\rho^-$,
measured its branching fraction
${\cal B}=(25^{+7+5}_{-6-6})\times 10^{-6}$,
and determined the longitudinal polarization fraction
${f_L}=0.98^{+0.02}_{-0.08}\pm 0.03$. 
Our quantitative estimates of penguin contributions in 
$B^0\to\rho^+\rho^-$ decays 
and the dominance of the $\CP$-even longitudinal 
polarization make this decay a promising channel for the 
measurement of the CKM angle $\alpha$.


We are grateful for the excellent luminosity and machine conditions
provided by our \pep2\ colleagues, 
and for the substantial dedicated effort from
the computing organizations that support \babar.
The collaborating institutions wish to thank 
SLAC for its support and kind hospitality. 
This work is supported by
DOE
and NSF (USA),
NSERC (Canada),
IHEP (China),
CEA and
CNRS-IN2P3
(France),
BMBF and DFG
(Germany),
INFN (Italy),
FOM (The Netherlands),
NFR (Norway),
MIST (Russia), and
PPARC (United Kingdom). 
Individuals have received support from the 
A.~P.~Sloan Foundation, 
Research Corporation,
and Alexander von Humboldt Foundation.


\bibliographystyle{h-physrev2-original}   %

\begin{thebibliography}{99}
\bibitem{chargeconj}
We use the name of one member of a charge conjugate pair
to refer to both unless explicitly stated otherwise.
\end{thebibliography}

\end{document}